\documentclass[10pt,a4paper,aps,showpacs,nofootinbib,
							amssymb,pra,floatfix,reprint,showkeys,superscriptaddress,
							]{revtex4-1}

\usepackage[tbtags,intlimits]{amsmath}

\usepackage{slashed}
\usepackage{xcolor}
\usepackage{hyperref}
\usepackage[squaren]{SIunits}

\usepackage[pdftex]{graphicx}
\usepackage{grffile} 

\usepackage{bbold}

\renewcommand{\mathbf}{\boldsymbol}
\renewcommand{\d}{  {\mathrm d}   }

\newcommand{\FAM}{F}

\begin{document}
\title{Structured x-ray beams from twisted electrons by inverse Compton scattering {of laser light}}
\author{D.~Seipt}
\email{d.seipt@gsi.de}

\affiliation{Helmholtz-Institut Jena, Fr{\"o}belstieg 3, 07743 Jena, Germany}

\author{A.~Surzhykov}
\affiliation{Helmholtz-Institut Jena, Fr{\"o}belstieg 3, 07743 Jena, Germany}

\author{S.~Fritzsche}
\affiliation{Helmholtz-Institut Jena, Fr{\"o}belstieg 3, 07743 Jena, Germany}
\affiliation{Universit\"at Jena, Institut f\"ur Theoretische Physik, 07743 Jena, Germany}


\begin{abstract}
{
The inverse Compton scattering of  laser light on high-energetic twisted electrons is
investigated with the aim to construct spatially structured x-ray beams. In particular, we analyze how the
properties of the twisted electrons, such as the topological charge and aperture angle of the electron Bessel beam,
affects the energy and angular distribution of scattered x-rays.}
We show that with suitably chosen initial twisted electron states
one can synthesize tailor-made x-ray beam profiles
with a well-defined spatial structure,
in a way not possible with ordinary plane-wave electron beams.
\end{abstract}

\maketitle

\section{Introduction}
When laser light collides with a beam of high-energy electrons
the photons are scattered predominantly
into a narrow angular region
and their frequency is up-shifted.
This is denoted as \emph{inverse} Compton scattering
and has been found an important x-ray source,
especially when using laser accelerated electrons
in order to realize an all-optical set-up
\cite{Phuoc:NatPhot2012,Luo:APL2013}.
The x-ray energy can be tuned from 
XUV \cite{Kazuyuki:JMicroNanolith2012},
hard x-rays \cite{Schoenlein:Science1996,Chouffani:NIMA2002,Jochmann:PRL2013}
up to $\gamma$ rays \cite{Chen:PRL2013}
by adjusting the electron energy, making it a versatile tool
for imaging the structure and dynamics of matter \cite{Schoenlein:Science1996,Luo:APL2013},
for medical \cite{Ikeura:APL2008}, biological \cite{Chouffani:NIMA2002} and nuclear physics applications \cite{Chen:PRL2013},
and for exploring strong field physics
\cite{Bula:PRL1996,Burke:PRL1997,DiPiazza:RevModPhys2012}.

While plane-wave electrons were used in most previous studies, new types of
electron vortex beams, called twisted electrons, can now be generated.
These twisted electrons
are characterized by the projection of orbital angular momentum $m$
onto the beam's propagation axis, which is the topological charge of the vortex.
Electron vortex beams were first predicted theoretically in 2007 \cite{Bliokh:PRL2007}
in analogy to optical vortex beams and
have been realized experimentally later on
with $m \sim 100\hbar$ and kinetic energies of several hundred keV
\cite{Verbeeck:Nature2010,Uchida:Nature2010,McMorran:Science2011}.
Twisted electrons have been used to
reveal information on the chiral and magnetic structure of
materials \cite{Lloyd:PRL2012,Yuan:PRA2013},
even down to the atomic scale \cite{Verbeeck:APL2011,Rusz:PRL2013}.
Moreover, they feature
spin-orbit-coupling \cite{Bliokh:PRL2011,Karlovets:PRA2012} and a large orbital angular momentum induced contribution to the electron magnetic moment \cite{Bliokh:PRL2011} that, e.g., modifies
polarization radiation
\cite{Ivanov:PRL2013,Ivanov:PRA2013}.
In addition, the propagation of twisted electron beams in
external electric or magnetic fields allows to study the vacuum
Faraday effect \cite{Greenshields:NJP2012},
Larmor and Guoy rotation \cite{Guzzinati:PRL2013}, and
fundamental properties of quantized electron states in magnetic fields \cite{Bliokh:PRX2012}.

\begin{figure}[!t]
\includegraphics[width=1.\columnwidth]{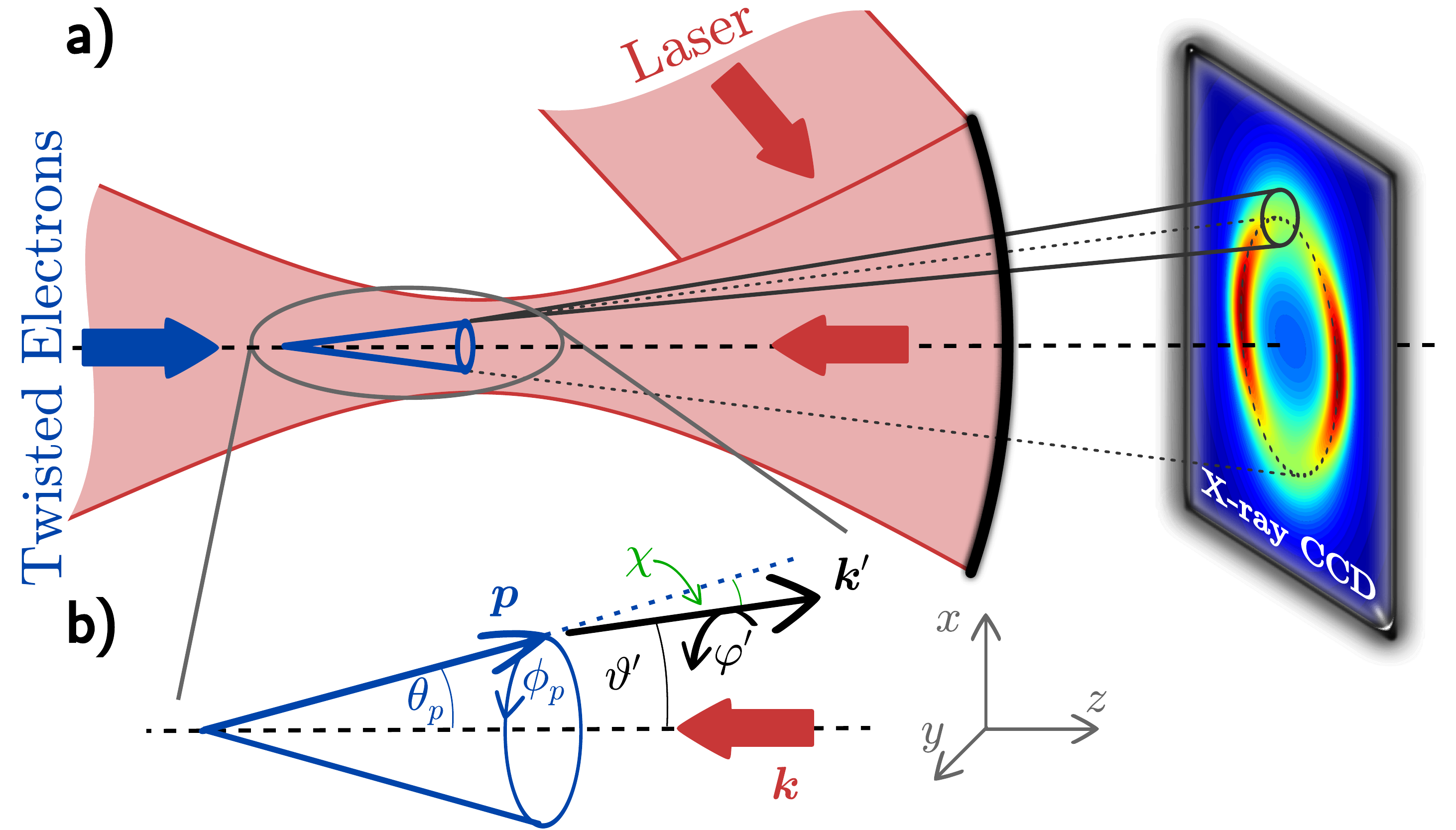}
\vspace*{-0.4cm}
\caption{(Color online) Compton scattering of laser light on a beam of twisted electrons helps
to generate x-rays with tailor-made spatial intensity distributions.
a) The light from an optical laser is focused by a parabolic mirror and collides head-on 
(under $\unit{180}{\degree}$) with electrons from a twisted beam (blue cone). The photons are mainly
backscattered, frequency up-shifted to x-ray energies and recorded on a CCD screen downstream the beam of electrons. 
Plane-wave laser light is assumed with an extent in the focal plane much larger than the waist of the electron beam.
b) Definition of coordinates and angles that are used to characterize the propagation of electrons
and photons in Compton scattering.
\vspace*{-0.4cm}}
\label{fig:1}
\end{figure}

In this paper, we consider the inverse Compton scattering of laser light on a beam of
twisted electrons. In particular, we here explore
the energy- and angular distribution of the 
emitted x-rays and compare them with those as obtained with ordinary plane-wave electrons.
{This analysis is complementary to previous investigations \cite{Jentschura:PRL2011,Jentschura:EPJC2011},
in which the Compton scattering of twisted photons on plane-wave electrons was studied
for the transfer of OAM to the backscattered photons.
For the present set-up with twisted electrons,
we find distinct features for the spatial and energy structure of the emitted x-rays that are 
not available for plane-wave electrons.}
We investigate the possibilities to utilize the degrees of freedom of the twisted electrons,
in particular their aperture angle and topological vortex charge,
to shape the spatial intensity distribution of emitted x-rays.
We show that a much better control of the spatial distribution of x-rays can be achieved by using proper beams of twisted electrons, instead of plane-wave electrons, in inverse Compton scattering x-ray sources.

Figure~\ref{fig:1} outlines the basic set-up for the inverse Compton scattering: A relativistic beam of
twisted electrons with total energy $E$ propagating along the $z$-axis collides \textit{head-on},
i.e.~under $\unit{180}{\degree}$, with optical laser light of frequency $\omega$.
For relativistic electrons,
of course, the photons are dominantly backscattered and shifted towards x-ray energies, and can thus 
be recorded quite easily downstream the beam of electrons on a CCD screen.
Behind the interaction point, the electrons should be bent away from the x-ray beam by a static
magnetic field \cite{Powers:NatPhoton2014}.

Our paper is organized as follows:
In Section \ref{sect:theory} we present the theoretical background, starting from a concise reminder
of the case of plane wave electrons and then generalizing to twisted electron beams.
In Section \ref{sect:results} we present the results of our calculation, starting with the energy distribution
of the scattered x-rays at fixed scattering angle and then going to the energy-integrated angular distributions. We finally discuss the scattering with superpositions of twisted electrons.
In Section \ref{sect:summary} we summarize and conclude.

\section{Theoretical Background}
\label{sect:theory}
\subsection{Cross Section for Plane Wave Electrons}

Before we investigate Compton scattering of light on twisted
electrons, let us briefly recall the usual case of
Compton scattering on \textit{plane-wave} electrons, for which the cross section
is readily derived by squaring the plane-wave S matrix \cite{book:Landau4}
\begin{align}
   S_{\rm pw}  &=  (2\pi)^4 i \,\delta^{\,(4)}(p+k-p'-k') \: \mathcal M
\label{eq:SPW}
\end{align}
with the scattering amplitude 
$  \mathcal M   =  4\pi \alpha \, \epsilon'^*_\mu \, \epsilon_\nu  \:  Q^{\mu\nu} $
and the Compton tensor
\begin{align}
   Q^{\mu\nu}  &=  
    \bar u_{p'\lambda'}\Big(
    	\gamma^\mu G(p+k) \gamma^\nu \:+\: \gamma^\nu G(p-k') \gamma^\mu
   \Big)u_{p\lambda}   \,.
\end{align}
In these formulas, 
$k=(\omega,\mathbf k)$ and $\epsilon$ as well as
$k'=(\omega',\mathbf k')$ and $\epsilon'$
are the four-momenta and polarization vectors of the incident and scattered photons,
while $p=(E,\mathbf p)$ and $p'=(E',\mathbf p')$ denote the 
four-momenta of the initial and final electrons.
The $\delta$-function in Eq.~\eqref{eq:SPW} ensures the 
conservation of the four-momenta
during the scattering.
Moreover, $\alpha$ is the fine structure constant, $\gamma^\mu$ denotes the Dirac matrices and $G$ is the electron propagator, and we use natural units with $\hbar=c=1$.
Note that the Dirac bi-spinors $u_{p\lambda}$ 
refer to \textit{plane-wave} electrons with momentum $p$ and helicity $\lambda$.%

The
angle-differential cross section for the Compton scattering on unpolarized plane-wave electrons is given 
by \cite{book:Landau4}
\begin{align}
\frac{\d \sigma_{\rm pw}}{\d\Omega'} &=  \frac{1}{128\pi^2}
\frac{\omega'(\vartheta')^2}{(k\cdot p)^2}
 \sum_{\lambda,\lambda',\epsilon'} |\mathcal M|^2 
= \alpha^2 \frac{\omega'(\vartheta')^2}{(k\cdot p)^2}  W\,,
\label{eq:cross_section_PW}
\end{align}
provided that the polarization of the scattered photons
remains unobserved.
In this expression, the sum 
over the squared scattering amplitude also determines the angular emission pattern,
\begin{align}
   W &=
   1 \:-\: \frac{m_e^2}{(k'\cdot p)^2} \, 
   \left| \epsilon\cdot k' \:-\:  \epsilon\cdot p \frac{k'\cdot k}{p\cdot k} \right|^2 \,,
\label{eq:def_W}
\end{align}
which depends on the polarization of the incident laser light. For a linearly-polarized laser,
for example, the emission pattern has a purely dipolar shape if all the terms including the electron recoil, 
$\gamma\omega/m_e \ll 1$, are omitted here and in the following
as is appropriate for optical laser frequencies and not too large electron energies.
In addition, the frequency of the scattered photons in
\eqref{eq:cross_section_PW}
\begin{align}
\omega'(\vartheta')  &=  \frac{4\gamma^2 \omega}{1+\gamma^2 \vartheta'^2} \,,
\label{eq:omega_approx}
\end{align}
depends on the polar angle $\vartheta'$ w.r.t.\ the beam of incoming electrons, the $z$-axis,
and is just given by the laser frequency $\omega$ as well as the Lorentz factor $\gamma = E/m_e\gg1$ of the 
electrons. In fact, Eq.~\eqref{eq:omega_approx} is crucial in order to understand the properties of the generated
x-rays: As seen from this expression, the laser frequency $\omega$ is Doppler shifted up to the
maximum frequency $\omega'_{\rm max}=4\gamma^2 \omega$ for $\vartheta'=0$; and $\omega'(\vartheta')$ decreases
rapidly as $\vartheta'$ is increased.
Moreover, since the cross section \eqref{eq:cross_section_PW} is proportional to $\omega'(\vartheta')^2$, 
most of the radiation is emitted into a narrow cone, centered around the electron momentum $\mathbf p$, and with
an angular divergence $\vartheta'\sim 1/\gamma$.

\subsection{Theoretical Description of Twisted Electron Beams}

Equations~\eqref{eq:cross_section_PW}--\eqref{eq:omega_approx} all refer to the
Compton scattering of laser photons on plane-wave electrons which are
characterized by the linear momentum $\mathbf p$.
However,
electron beams
can be prepared also in
twisted states, known as electron vortex beams \cite{Verbeeck:Nature2010,Uchida:Nature2010,McMorran:Science2011},
and characterized by the linear momentum component $p_z$ and the projection of orbital angular 
momentum (OAM) onto the propagation axis $\ell_z=m$, which
is the topological charge of the vortex \cite{Bliokh:PRL2007}.
For Bessel beams of twisted electrons, especially,
only the absolute value of transverse momentum has
a unique value $|\mathbf p_\perp|= \varkappa$ while its direction remains undefined, and
the wave function can be written as \cite{Ivanov:PRA2011,Karlovets:PRA2012}
\begin{align}
\psi_{\rm tw}(x) &= \int \! \frac{\d^2 \mathbf p_\perp}{(2\pi)^2} \, a_{\varkappa m}(\mathbf p_\perp) 
e^{-ip\cdot x} u_{p\lambda}
\label{eq:twisted_electron}
\end{align}
with the amplitudes
\begin{align}
a_{\varkappa m }(\mathbf p_\perp ) = \sqrt{\frac{2\pi}{\varkappa}} (-i)^m e^{im\phi_p} \delta(|\mathbf p_\perp|-\varkappa) \,.
\label{eq:amplitudes}
\end{align}
As seen from Eq.~\eqref{eq:twisted_electron}, the twisted Bessel beams are
coherent superpositions of plane-wave states $\psi_{\rm pw}(x) = e^{-ip\cdot x} u_{p\lambda}$
with momentum vectors
$\mathbf p(\phi_p) = (\varkappa \cos \phi_p, \varkappa \sin \phi_p,p_z)$
that are distributed
uniformly over the azimuthal angle $\phi_p$ and
that 
form the surface of a cone with an aperture $\theta_p = \arctan \varkappa/p_z$
(see Fig.~\ref{fig:1}~b).

\subsection{Cross Section for Twisted Electrons}

With the description of twisted electrons at hand we are ready to
study Compton scattering of laser light
with the initial electrons in a \emph{twisted state}.
To calculate the cross section we have to
square the twisted S matrix \cite{Jentschura:PRL2011,Ivanov:PRD2011}
\begin{align}
S_{\rm tw}
&= \int \! \frac{\d^2 \mathbf p_\perp}{(2\pi)^2} \, a_{\varkappa m}(\mathbf p_\perp) 
S_{\rm pw}(\mathbf p_\perp)\,,
\label{eq:STW}
\end{align}
which is the superposition of plane-wave S matrices, Eq.~\eqref{eq:SPW}, 
with the same amplitudes $a_{\varkappa m}(\mathbf p_\perp)$ as in \eqref{eq:twisted_electron}.
{Owing to the conservation of the linear momentum, the squared S matrix simplifies to \cite{Ivanov:PRD2011}
%
\begin{align}
|S_{\rm tw}|^2
&= (2\pi)^4 \int \! \d ^2 \mathbf p_\perp  \d^2 \bar{\mathbf p}_\perp \delta^{(4)}(p+k-p'-k')  \nonumber \\
& \quad \times \delta^{(4)}(\bar p - p ) a_{\varkappa m }^*(\bar{\mathbf p}_\perp) a_{\varkappa m}(\mathbf p_\perp) \mathcal M^*(\bar p) \mathcal M(p)  \nonumber \\
&\propto
  (2\pi)^4  \int \! \d ^2 \mathbf p_\perp \: \delta^{(4)}(p+k-p'-k') \nonumber \\
& \quad \times |a_{\varkappa m}(\mathbf p_\perp)|^2
|\mathcal M(p)|^2
\end{align}
%
with $\bar p = (E_{\bar{\mathbf p}},\bar{\mathbf p})$ and $\bar{\mathbf p} = (\bar{\mathbf p}_\perp,p_z)$. In this expression only the
absolute square of the amplitude \eqref{eq:amplitudes} appears,
$|a_{\varkappa m}(\mathbf p_\perp)|^2
= \frac{2\pi}{\varkappa} [\delta(|\mathbf p_\perp|-\varkappa)]^2$,
while the dependence on the OAM value $m$ drops out.
The square of the radial delta-function can be
re-written as $[\delta(|\mathbf p_\perp|-\varkappa)]^2\to \delta(|\mathbf p_\perp|-\varkappa) \frac{R}{\pi}$ for a large but finite cylindrical normalization volume with radius $R$ for the
twisted electron wave function. In the cross section, in fact, this factor $R$ cancels with the one that occurs in the normalization of the 
initial twisted state \cite{Ivanov:PRD2011,Jentschura:EPJC2011}.}
{Moreover, in the definition of a cross section for twisted particles one has to take into account
that the flux of incident particles in a twisted wave depends on the lateral position. It has been proposed to define an average cross section by averaging the plane wave flux factor $I_{\rm pw}$ over all plane wave components as $I_{\rm tw} = \int \frac{\d \phi_p}{2\pi} {I_{\rm pw}(p)}$ \cite{Ivanov:PRD2011}.
For head-on collisions as considered in the present case, the averaged flux factor coincides however with the plane wave flux factor $I_{\rm tw} = I_{\rm pw} = k\cdot p$.}

Since the squared amplitude is
independent of the OAM value $m$,
the energy and angle-differential cross section for Compton scattering of laser light on unpolarized \emph{twisted} electrons
\begin{align}
\frac{\d \sigma_{\rm tw}}{\d\Omega'\d \omega'}
		&= 
	\int \! \frac{\d \phi_p}{2\pi} \: 
	\delta  \big( \omega' -\omega'\mathbf(\chi (\phi_p) \mathbf) \big) 
		  \frac{ \alpha^2\omega'^2 }{(k\cdot p)^2} \: W\mathbf( p(\phi_p) \mathbf ) \,,
\label{eq:cross_section_TW_master}
\end{align}
only depends on the momentum cone aperture $\theta_p$
but not on the vortex charge $m$.
Because the plane-wave component's momenta $\mathbf p(\phi_p)$ are lying on
a cone surface, both the plane-wave emission pattern $W$ and the
frequency of the scattered photons $\omega'\mathbf(\chi(\phi_p) \mathbf)$ now depend on the electron azimuthal angle $\phi_p$.
In particular,
\begin{align}
\omega'\mathbf(\chi(\phi_p)\mathbf) = \frac{4\gamma^2 \omega}{1+\gamma^2 \chi^2(\phi_p)}
\label{eq:frequency_chi}
\end{align}
resembles Eq.~\eqref{eq:omega_approx}, but with
the photon scattering angle $\vartheta'$ replaced by an effective scattering angle
$\chi(\phi_p)$:
the angle between $\mathbf p(\phi_p)$ and the momentum $\mathbf k'$ of the scattered photon.
A simple expression for the effective scattering angle $\chi(\phi_p)$
can be obtained in the regime of small cone aperture $\theta_p\ll1$ and photon scattering angles
$\vartheta'\ll1$, where (cf.~Fig.~\ref{fig:1} b)
\begin{align}
\chi^2(\phi_p) = \theta_p^2+\vartheta'^2 
				- 2 \theta_p \vartheta' \cos (\phi_p-\varphi') \,.
				\label{eq:effective_angle}
\end{align}

\section{Discussion of Results}
\label{sect:results}
{
\subsection{The Energy Distribution of Scattered Photons}

For Compton scattering of (plane-wave) laser light on twisted electron beams the effective scattering angle $\chi(\phi_p)$ depends on the azimuthal angle $\phi_p$ of the plane-wave components.
This effective scattering angle in turn determines the frequency of the scattered photons $\omega'(\chi)$ according to Eq.~(10).
Since we have to integrate over $\phi_p$ in Eq.~(9), we get a distribution of frequencies $\omega'$ of the scattered photons for fixed scattering angle $\vartheta'$, in contrast to the case of plane-wave electrons where the frequency of the scattered photons is determined for a given angle $\vartheta'$, see Eq.~(5).
Figure \ref{fig:energy_differential} a) displays
the range of allowed frequencies $\omega'_-(\vartheta')\leq \omega'\leq \omega'_+(\vartheta')$ as shaded area, and where the lower and upper bound of this frequency range is given by
\begin{align}
\omega'_\pm(\vartheta') &= \frac{4\gamma^2\omega}{1+\gamma^2(\theta_p\mp \vartheta')^2} \,.
\end{align}
While the upper bound $\omega'_+(\vartheta')$ of the spectrum arises from the plane-wave
component that is maximally aligned with the scattered photon momentum $\mathbf k'$, i.e.~$\phi_p=\varphi'$, with the smallest effective scattering angle
$\min_{\phi_p} \chi(\phi_p) = |\theta_p-\vartheta'|$, the 
lower bound $\omega'_-(\vartheta')$ is caused by the least aligned plane-wave component with $\phi_p=\varphi'+\pi$ and
the largest effective scattering angle $\max_{\phi_p} \chi(\phi_p) = \theta_p+\vartheta'$.
For comparison, Figure~\ref{fig:energy_differential} a) also shows the plane-wave energy angle correlation (5) as dashed curve (pw).
Although the maximum frequency $\omega'_{\rm max} = 4\gamma^2 \omega$ is equal for both plane-wave and twisted electrons, its position shifts from $\vartheta'=0$ for
plane-wave electrons
to $\vartheta'=\theta_p$ for twisted electrons, where the
momentum $\mathbf p$ of one particular plane-wave component is supposed to be
collinear with $\mathbf k'$, i.e.~$\chi=0$.

To calculate the distribution of scattered frequencies we
perform the integration over $\phi_p$ in (9) and use the
identity for the $\delta$-function
\begin{align}
\delta(f(x))&=
\sum_{x_i} \frac{\delta(x-x_i)}{|\d f(x_i)/\d x|} \,,
\end{align}
where the sum is over all zeros $x_i$ of $f(x)$.
Then, the energy and angle-differential cross section
\begin{align}
\frac{\d \sigma_{\rm tw}}{\d \Omega'\d\omega'} 
& =  \frac{\alpha^2}{2\pi} \left( \frac{\omega'}{k\cdot p} \right)^2 \frac{W(\varphi' + \phi_p^\star) + W( \varphi' - \phi_p^\star)}{\left| \frac{\d \omega'(\chi)}{\d \phi_p}\right|_{\varphi' + \phi_p^\star }} 
\label{eq:triple_differential}  
\end{align}
is expressed as the incoherent superposition of contributions from two particular
plane-wave components with azimuthal angles $\phi_p = \varphi'\pm \phi_p^\star(\omega')$
with
\begin{align}
\phi_p^\star(\omega') &= \arccos 
\frac{\omega' (1+\gamma^2\theta_p^2+\gamma^2\vartheta'^2) -4\gamma^2 \omega}{2\gamma^2 \theta_p\vartheta' \omega'} \,.
\end{align}
{
The coherence of the final state particles is hidden
because we project the final states onto plane-waves with different linear momenta
that do not interfere \cite{Ivanov:PRD2011}.
In particular, the two plane-wave contributions in Eq.~\eqref{eq:triple_differential}
have different final electron momenta $p'$
and they are added incoherently on the level of the squared amplitudes.
The plane-wave final states are a convenient choice
since we study the energy and angular distributions of the scattered x-rays,
which are observed by a detector that measures the photons'
linear momentum but not their orbital angular momentum \cite{Ivanov:PRD2011}.
To access the coherence of the final states, e.g.~the orbital angular momentum of the scattered photons,
one has to project the final particles onto a basis of twisted states, as was done e.g.~in \cite{Jentschura:EPJC2011}.
}

Numerical results for the energy and angle-differential cross section \eqref{eq:triple_differential}
are
displayed in Fig.~\ref{fig:energy_differential} b) for selected values of the photon scattering angle $\vartheta'$.
As seen in Fig.~\ref{fig:energy_differential} b), the energy-spectrum strongly peaks at the upper bound, which means that most of the scattered photons have
an energy close to $\omega'_+(\vartheta')$.
To account for a finite detector resolution, we have averaged the spectra in 
Fig.~\ref{fig:energy_differential} with a resolution of $\Delta \omega' = \unit{10}{\electronvolt}$,
which also removes the integrable divergence in Eq.~\eqref{eq:triple_differential} due to the zero derivative $|\d \omega'(\chi) / \d \phi_p |$ at both the upper and lower bounds of the energy spectrum.

\begin{figure}[!t]
\begin{center}
\includegraphics[width=\columnwidth]{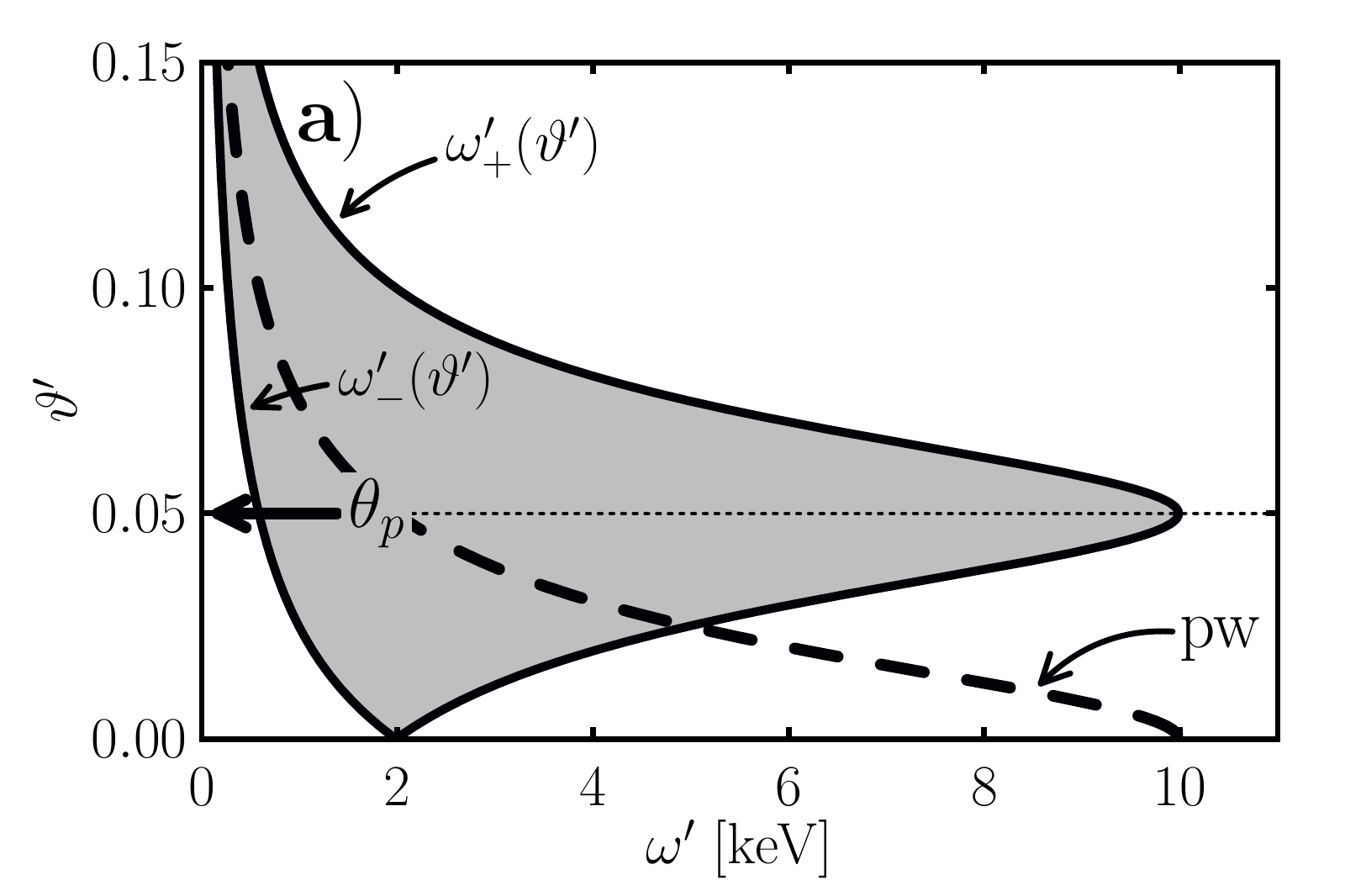}
\includegraphics[width=\columnwidth]{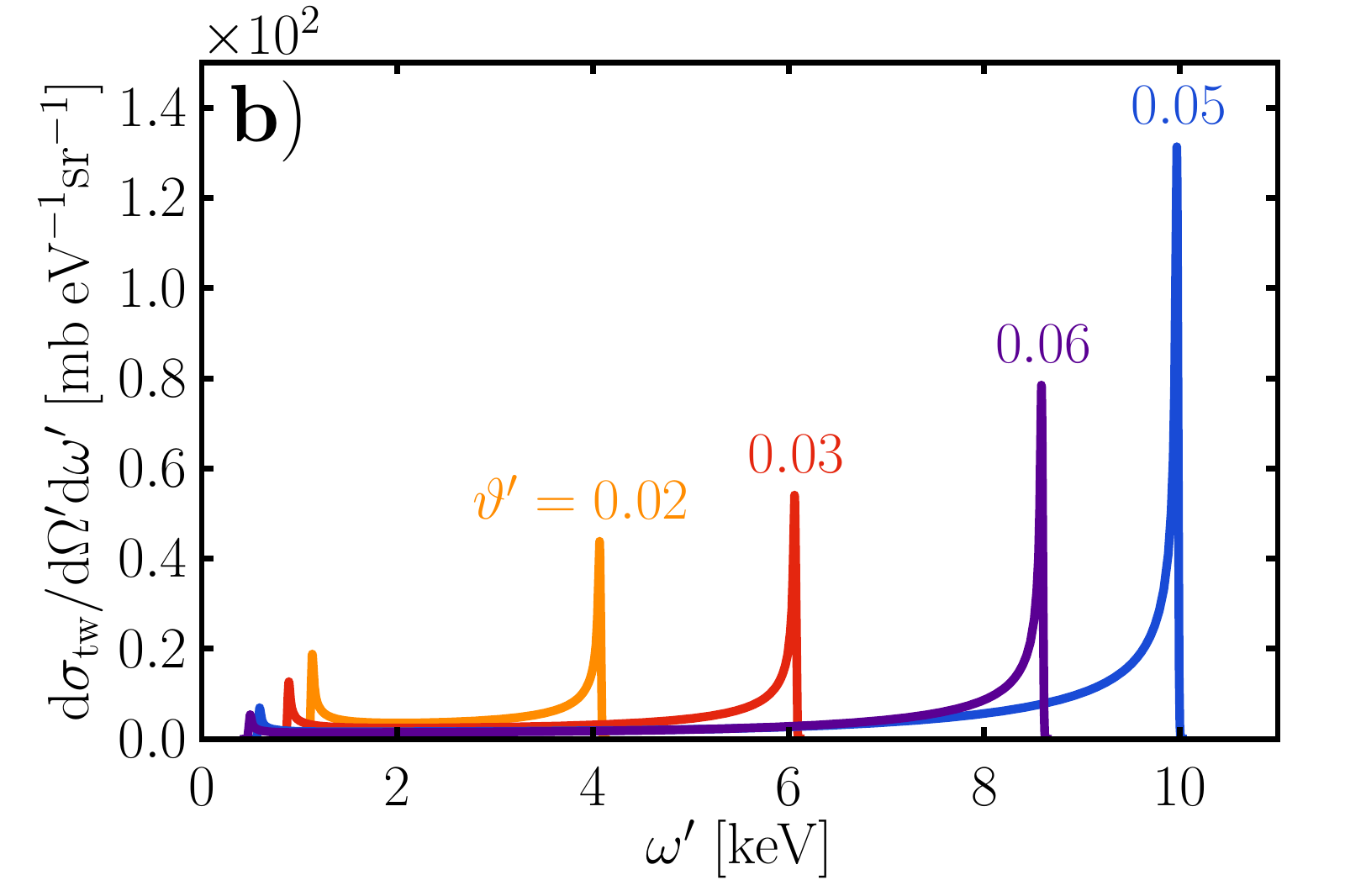}
\end{center}
\caption{{Energy distribution of scattered photons for Compton scattering of circularly polarized laser light with frequency $\omega=\unit{1.55}{\electronvolt}$ on twisted electrons with initial energy $E= \unit{20.5}{\mega\electronvolt}$, i.e.~$\gamma=40$, and
for a cone aperture $\theta_p=0.05$.
Panel (a) displays the allowed energies $\omega'_-(\vartheta') \leq \omega' \leq \omega_+(\vartheta')$
of the scattered photons (shaded area) for the predominant
scattering angles $0 \,\le\, \vartheta' \,\le\, 0.15$. For comparison, the dashed curve gives the
known energy-angle correlation for plane-wave electrons, cf.~Eq.~(5).
Panel (b): Energy and angle-differential cross section $\d \sigma_{\rm tw}/\d \Omega'\d \omega'$ for Compton scattering of plane-wave photons on twisted electrons as function of
the energy
$\omega'$ of the scattered photons,
and for selected values of the photon scattering angle $\vartheta'$.}
}
\label{fig:energy_differential}
\end{figure}

}

\subsection{Angular Distributions}

To understand the angular distribution of the scattered photons,
independent of their particular energy,
we integrate the cross section in \eqref{eq:cross_section_TW_master} over $\d \omega'$ and
obtain the angle-differential cross section for Compton scattering on twisted electrons
\begin{align}
\frac{\d\sigma_{\rm tw}}{\d \Omega'} &= 
		 \int \! \d \omega' \, \frac{\d \sigma_{\rm tw}}{\d \Omega' \d\omega' } = 
		 \int \! \frac{\d \phi_p}{2\pi} \, \frac{\d \sigma_{\rm pw} (\theta_p,\phi_p)}{\d \Omega'} \,,
\label{eq:cross_section_TW}
\end{align}
simply as average over the plane-wave cross section \eqref{eq:cross_section_PW} for
\emph{all plane-wave components} of the twisted electron beam.

\begin{figure}[!t]
\begin{center}
\includegraphics[width=0.50\columnwidth]{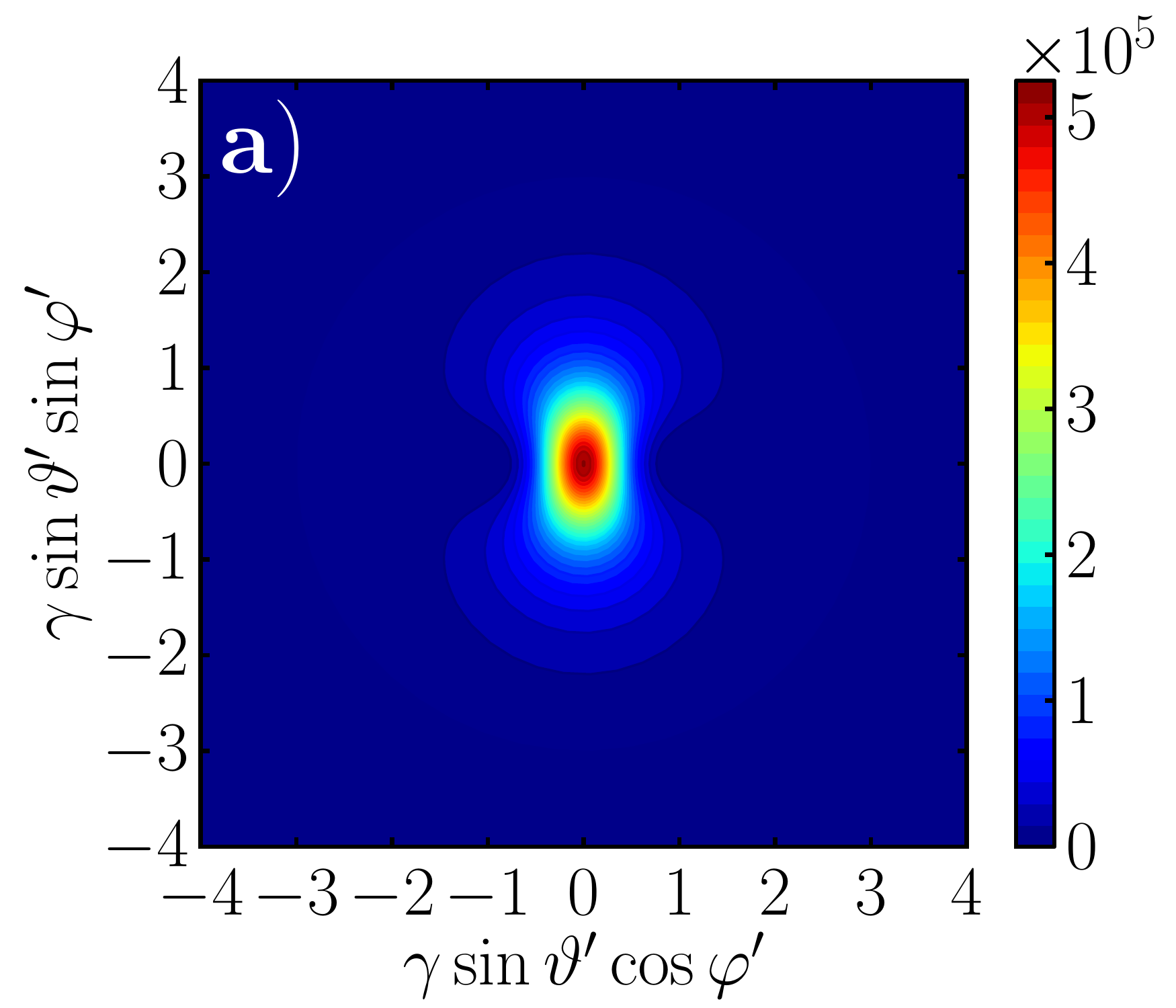}
\hspace*{-0.03\columnwidth}
\includegraphics[width=0.50\columnwidth]{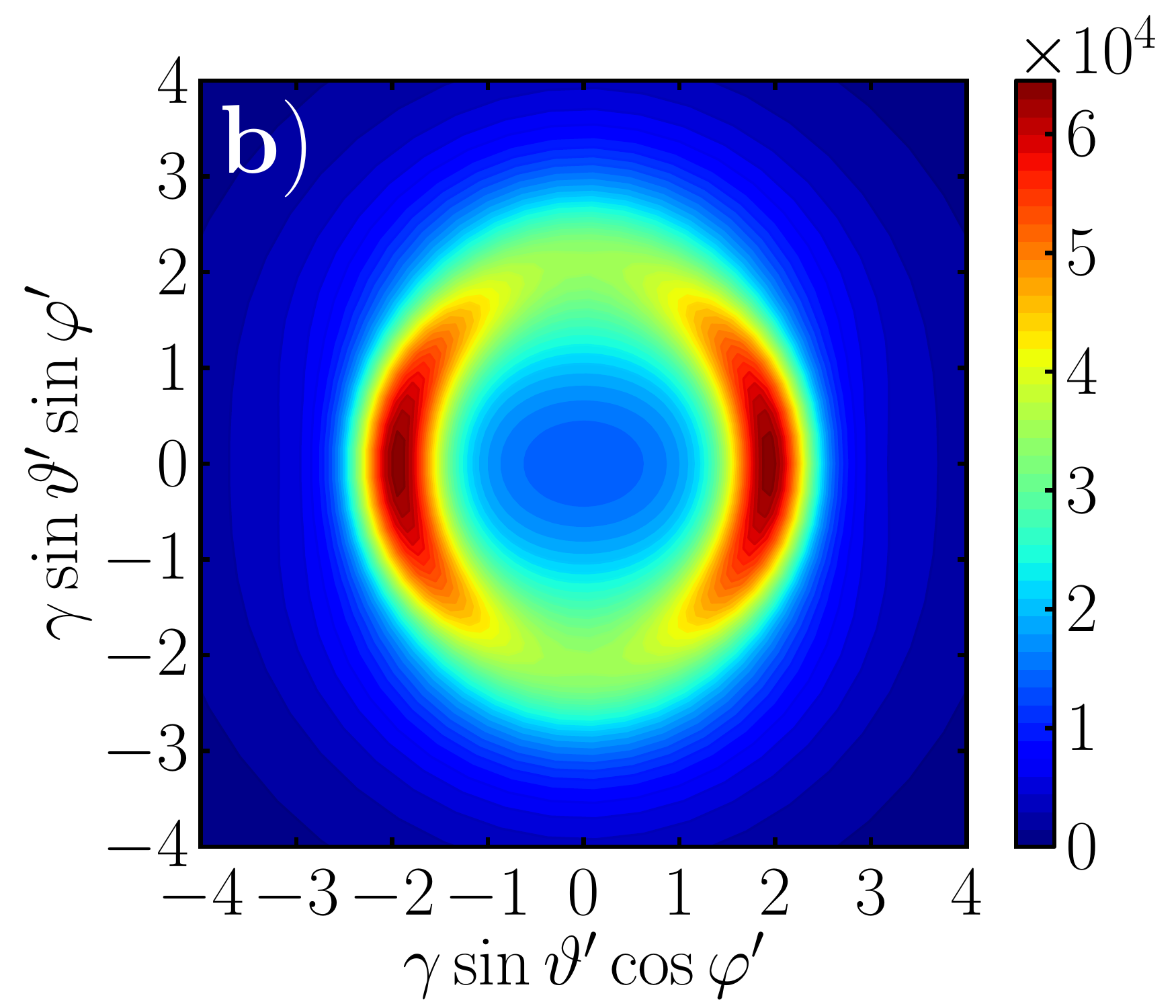}
\end{center}
\vspace*{-0.4cm}
\caption{(Color online) Angle-differential cross section for 
(a) plane-wave electrons and (b) a beam of twisted electrons with aperture angle $\theta_p=0.05$
and with initial energy $E= \unit{20.5}{\mega\electronvolt}$, i.e.~$\gamma=40$, in both cases.
The laser light is linearly polarized along the $x$-axis and has the frequency $\omega=\unit{1.55}{\electronvolt}$.
}
\label{fig:plane_twisted}
\end{figure}

Figure~\ref{fig:plane_twisted} shows
the angle-differential cross section \eqref{eq:cross_section_TW} as observed
downstream the initial electron beam, e.g.~on the screen of an x-ray CCD camera.
Results for plane-wave electrons with an intensity distribution of emitted photons around
$\vartheta'=0$ (a) 
are compared with those for twisted electrons (b).
For plane-wave electrons the cross section \eqref{eq:cross_section_PW}
has a peak at the photon scattering angle $\vartheta'=0$,
where the outgoing photon momentum $\mathbf k'$ is collinear
with the initial electron momentum $\mathbf p$, with an angular divergence of
$\vartheta'\sim 1/\gamma$ and a dipolar pattern due to the linear laser polarization.
For twisted electrons, in contrast, each of the plane-wave components produces
an analog pattern with the peak intensity at $\vartheta'=\theta_p$
and $\varphi'=\phi_p$.
The average over all $\phi_p$ in \eqref{eq:cross_section_TW} leads
to the formation of a hollow-cone of x-ray intensity, with a local minimum
on the beam axis, $\vartheta'=0$, and the highest intensity on a ring around the beam axis at
$\vartheta' \approx \theta_p$ {(In Fig.~\ref{fig:plane_twisted} b) this ring has a radius of $\gamma \sin \theta_p \approx 2$.)}, which just coincides with the angle where the
largest x-ray energies occur.

Due to the dipole shape of the plane-wave emission pattern the
intensity is distributed nonuniformly over the ring with
two clear maxima along the the laser polarization direction, i.e.~the $x$-axis.
For circular laser polarization the distributions in
Fig.~\ref{fig:plane_twisted} would be azimuthally symmetric.
The size of the ring of highest intensity can be adjusted by varying the cone aperture $\theta_p$, and
in the limit $\theta_p\to 0$ the cross section for twisted electrons coincides with the plane-wave
cross section.

\subsection{Angular Distributions for a Superposition of Twisted Electrons}

All x-rays spectra as obtained so far for the scattering
of laser light on twisted electrons
are sensitive to the aperture $\theta_p$ of the momentum cone but not to
their vortex charge $m$.
To further control the properties of the scattered photons, we wish to utilize
this \textit{additional} OAM degree of freedom and
explore how the scattering is affected by a superposition of twisted electrons
with two different values of the vortex charge.
Such superpositions have been analyzed, e.g.~for the observation of 
Guoy and Larmor rotations \cite{Guzzinati:PRL2013}.
A superposition of electrons with just two values of the vortex charge is described by Eq.~\eqref{eq:twisted_electron}
with the modified amplitude
\begin{align}
A_{\varkappa,m_1m_2}(\mathbf p_\perp) =\frac{1}{\sqrt2} \left[ 
     	     a_{\varkappa m_1}(\mathbf p_\perp) 
     	      +  e^{i\delta} a_{\varkappa m_2}(\mathbf p_\perp)
     	      \right] \,,
\label{eq:superposition}
\end{align}
but for which the plane-wave components are no longer distributed uniformly over the
azimuthal angle $\phi_p$.
This can be seen also from the square of the amplitudes
\begin{align}
|A_{\varkappa,m_1m_2}(\mathbf p_\perp)|^2 = 
\frac{2\pi}{\varkappa} [\delta(|\mathbf p_\perp|-\varkappa)]^2 \FAM (\phi_p)
\end{align}
where the azimuthal distribution
\begin{align}
\FAM (\phi_p) = 
	1 + \cos \left[ \Delta m \left( \phi_p-\frac{\pi}{2} \right) +\delta \right]\,.
	\label{eq:def_FAM}
\end{align}
depends on the difference of the vortex charges $\Delta m = m_2-m_1$.
The distribution \eqref{eq:def_FAM} and, hence, $\Delta m$ also determines
the angle-differential cross section
\begin{align}
\frac{\d \sigma_{\rm 2tw}}{\d \Omega'} =
		 \int \! \frac{\d \phi_p}{2\pi} 
		 \FAM (\phi_p)
		 \frac{\d \sigma_{\rm pw}(\theta_p,\phi_p)}{\d \Omega'} \,.
		 \label{eq:double_twisted}
\end{align}
for a superposition of two twisted electron states.
\begin{figure*}[!p]
\begin{center}
\includegraphics[width=1.8\columnwidth]{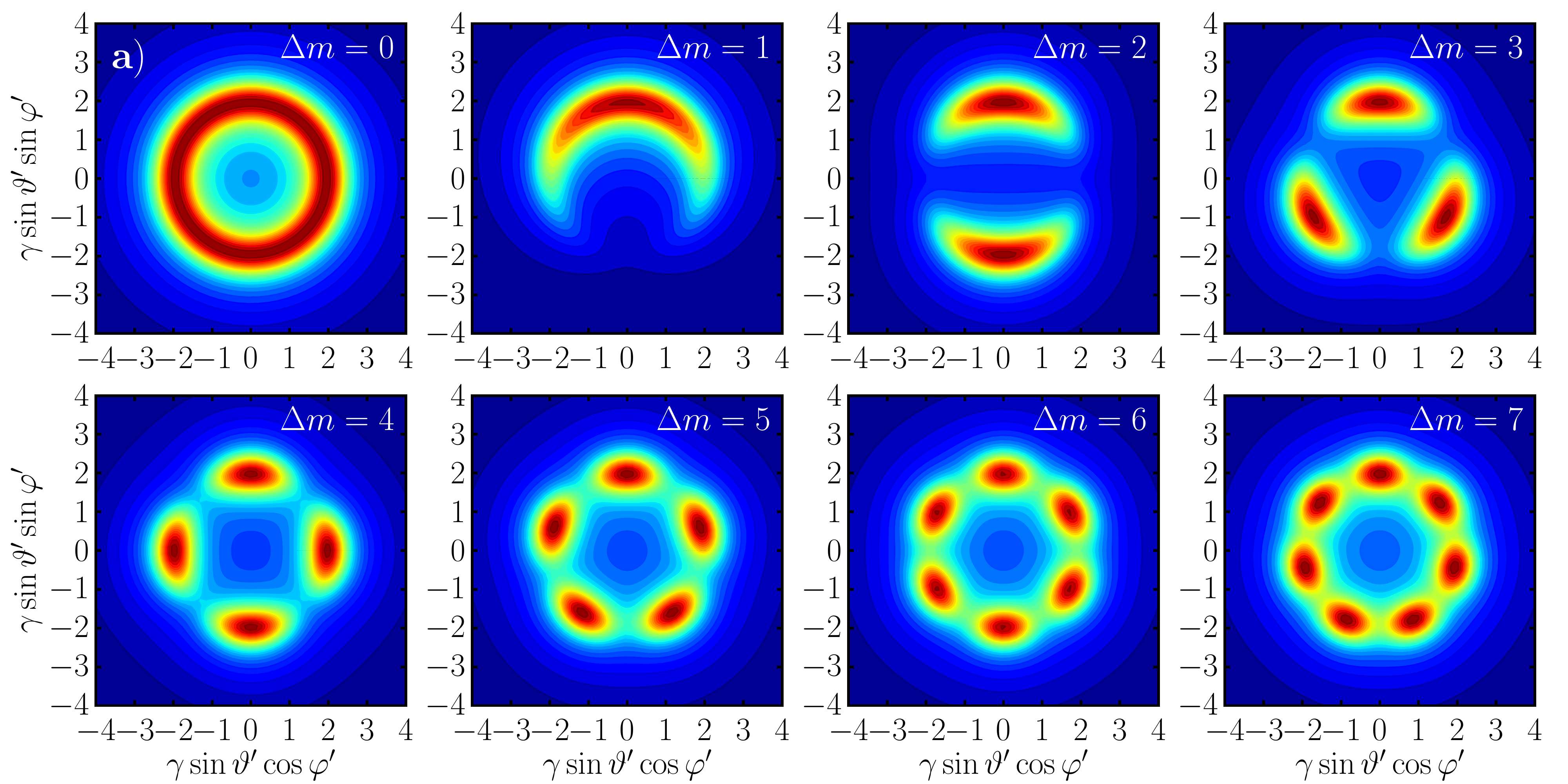}

\vspace*{1mm}
\includegraphics[width=1.8\columnwidth]{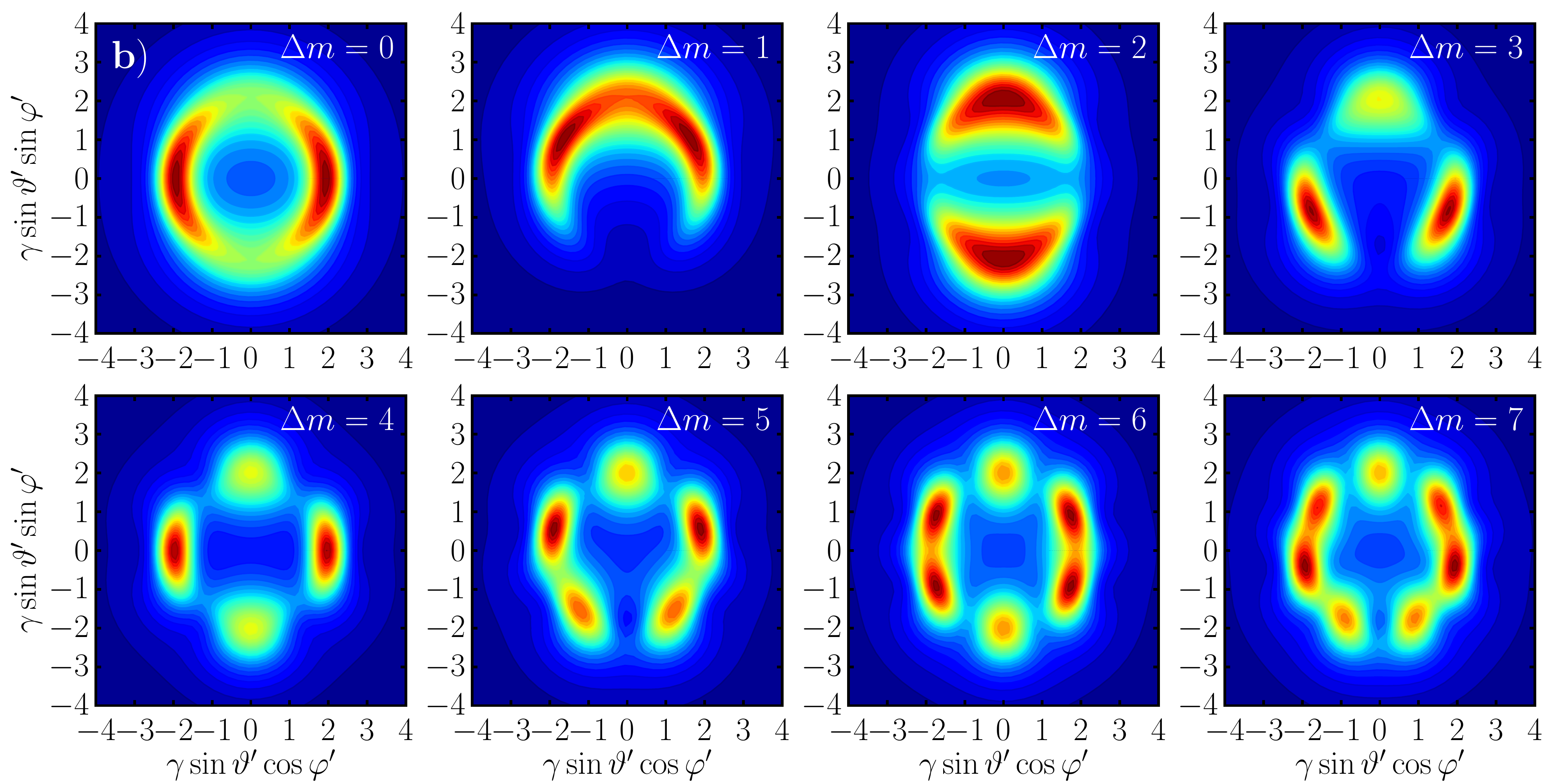}

\vspace*{1mm}
\includegraphics[width=1.8\columnwidth]{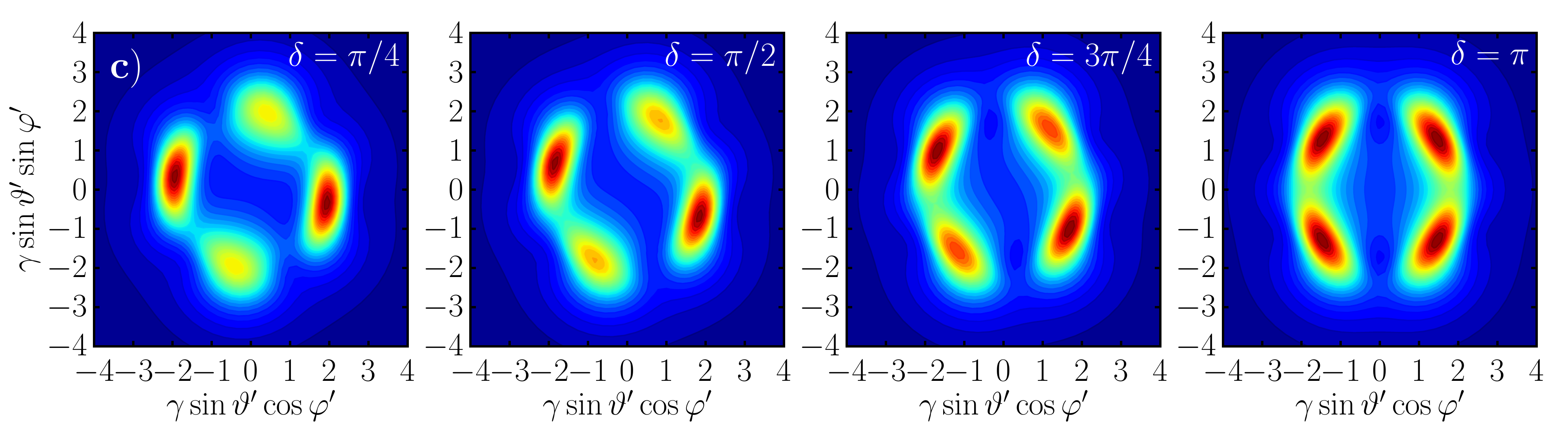}
\end{center}
\caption{(Color online) Angle-differential cross section \eqref{eq:double_twisted} for Compton scattering 
of plane-wave laser photons on a superposition of two equally intense beams of twisted electrons 
with given difference of the vortex charge $\Delta m$ and relative phase $\delta$, cf.~Eq.~\eqref{eq:superposition}. 
The scale of the $x$ and $y$ axes and the colors are the same as in Fig.~\ref{fig:plane_twisted}. Results 
are shown for selected values of $\Delta m$ and $\delta=0$ for (a) circularly-polarized laser light and 
(b) linearly-polarized light along the $x$-axis.
The panels (c) display the sensitivity of the cross section on
the relative phase $\delta$ for fixed $\Delta m=4$ and linear laser polarization.
The calculations were done for the same parameters as in Fig.~\ref{fig:plane_twisted}.}
\label{fig:azimuthal}
\end{figure*}

Equation \eqref{eq:double_twisted} represents a \emph{weighted average} of
the plane-wave cross sections
with weight $\FAM(\phi_p)$, which 
acts as a mask that modulates the cross section on top
of the hollow-cone shape of the x-ray beams discussed above.

The difference $\Delta m$ in the vortex charge of the electrons provides
us with a direct control of the spatial distribution of the emitted photons.
For example, Fig.~\ref{fig:azimuthal}~a) shows the distributions
for circularly polarized laser light with $|\Delta m|$ maxima of the x-ray intensity,
located at the photon azimuthal angles
\begin{align}
\varphi'_n = \frac{2\pi n}{\Delta m} + \frac{\pi}{2} - \frac{\delta}{\Delta m} 
\,, \quad n=0, 1,\ldots, |\Delta m|-1 \,,
\label{eq:phi_n}
\end{align}
where the outgoing photon momentum $\mathbf k'$ is collinear with a maximum-weight plane-wave component.
For circular laser polarization the $|\Delta m|$-fold symmetry of the patterns
in Fig.~\ref{fig:azimuthal}~a) directly reflects the symmetry of the azimuthal distribution $\FAM(\phi_p)$
of plane-wave components.
This symmetry not visible
for linear laser polarization in Fig.~\ref{fig:azimuthal}~b), where the mask $\FAM(\phi_p)$ acts on the
dipolar shape of the hollow-cone and leads to peaks of different intensity.
The effect of the relative phase $\delta$ is a rotation of the azimuthal mask $\FAM(\phi_p)$
by an angle $-\delta / \Delta m$ (cf.~Eq.~\eqref{eq:phi_n}).
For circular polarization this induces just a rotation
of the patterns by that angle; for linear polarization the patterns change due to the
inherent dipole distribution which is not rotated (see Fig.~\ref{fig:azimuthal} c).

\section{Summary and Conclusion}
\label{sect:summary}

{In summary, we investigated the energy and angular distribution of the scattered x-rays
in inverse Compton scattering of (plane-wave) laser light on twisted electrons.
In particular, we found a strongly peaked energy distribution for fixed scattering angle,
and with the maximum frequency occurring
at a finite scattering angle that is equal to the cone aperture of the twisted electron beam.
The angular distributions of the scattered x-rays are spatially structured with
distinct patterns.
}
We showed that the cone aperture and vortex charge
of twisted electrons provide a direct way to
control these spatial distributions of emitted x-rays,
beyond the possibilities of ordinary plane-wave electrons.

A key feature for the experimental verification of these effects is the availability
of high-energy twisted electron beams, which have not been produced so far.
A possible path to obtain high-energy twisted electrons is to expose
relativistic plane-wave
electron beams to the field of a magnetic monopole \cite{Dirac:ProcRoySoc1931}.
The feasibility of this scheme
has been demonstrated recently
\cite{Beche:NatPhys2014}.
{Moreover, the Bessel beams of twisted electrons described in our paper are considered as
an idealized scenario. A more realistic description would include the effects of a distribution
of transverse momenta, replacing in Eq.~\eqref{eq:amplitudes} $\delta(|\mathbf p_\perp|-\varkappa) \to f(\varkappa)$, with some narrow distribution of $\varkappa$ with width $\Delta \varkappa$
around some central value $\varkappa_0$. The patterns reported in Figs.~\ref{fig:plane_twisted} and
\ref{fig:azimuthal} will persist for $\Delta \varkappa \ll \varkappa_0$.}

{
In this paper we presented a theoretical study of the inverse Compton scattering
of laser light on relativistic twisted electron beams,
that furnishes a scheme to generate tailor-made structured
x-ray beams.}
%
%
%
%
The spatially structured x-rays could enhance the
capabilities
of x-ray imaging techniques
like x-ray microscopy or phase contrast imaging by providing a novel way for a
structured exposure of the samples.
In addition, detecting the distribution of scattered x-rays can help
to characterize the vortex charges of twisted electron beams.
Superpositions of more than two twisted electron states with different vortex charges and transverse momenta \cite{Ornigotti:2014} allow to generate much richer structures in
the angular distributions of the scattered photons, and
one can think about
reverse engineering the necessary twisted electron beams
for a desired distribution of x-rays.

%


\end{document}